\documentclass[12pt,aps,prd,superscriptaddress,showpacs,longbibliography,floatfix,nofootinbib]{revtex4-2}

\usepackage[utf8]{inputenc}
\usepackage{slashed}
\pdfoutput=1

\usepackage{color}
\usepackage{graphicx}   
\usepackage{bm}
\usepackage{amsmath}
\usepackage{amsfonts}
\usepackage{hyperref}

\def\be{\begin{equation}}
\def\ee{\end{equation}}
\def\ba{\begin{eqnarray}}
\def\ea{\end{eqnarray}}

\begin{document}

\title{A loophole in the proofs of asymptotic freedom and quantum triviality}

\author{Paul Romatschke}
\affiliation{Department of Physics, University of Colorado, Boulder, Colorado 80309, USA}
\affiliation{Center for Theory of Quantum Matter, University of Colorado, Boulder, Colorado 80309, USA}

\begin{abstract}
  In 1973, Coleman and Gross proved that in four dimensions, only non-abelian gauge theories can have asymptotic freedom. More recently, Aizenman and Duminil-Copin proved that four dimensional scalar field theories are quantum trivial in the continuum. Both of these proofs have a loophole, and it is the same loophole in both proofs: The proofs assume that the scalar self-coupling in the UV is positive definite. While this is a perfectly reasonable and classically very intuitive assumption, it is an assumption nevertheless. In this work, I show that the assumption of coupling positivity is violated in a concrete quantum field theory, the O(N) model, in the large N limit. Surprisingly, despite the classically nonsensical unbounded potential, the negative coupling has no pathological consequence for propagators, the free energy or cross sections. This suggests that interacting scalar field theories with asymptotic freedom in four dimensions are possible, despite long-held opinions to the contrary.
\end{abstract}

\maketitle

Classical physics is intuitive. In classical physics, a marble placed inside a bowl will always stay within the bowl, and never escape off to infinity. Similarly, if the bowl is turned upside-down, the marble place on its upper surface have the tendency to escape far away from the bowl.

In quantum physics, things are not so intuitive. Phenomena such as quantum tunneling or the two-slit experiment have no simple classical analogue, and it is only by learning to trust the mathematics that physicists were able to gain understanding of quantum phenomena.

The proofs of two important concepts in theoretical quantum field theory, namely the no-go proofs for asymptotic freedom and non-trivial interaction in four dimensional scalars in the continuum \cite{Coleman:1973sx,Aizenman:2019yuo}, are based on the assumption of a classically stable potential. The authors of Ref.~\cite{Aizenman:2019yuo} are very specific that their proof only applies to stable potentials, yet in the high-energy physics community, their proof often gets summarized as ``interacting scalar field theory in the continuum does not exist''. However, already in 1973, Symanzik \cite{Symanzik:1973hx,kleefeld2005kurt} suggested that classically unstable potentials may lead to well-defined quantum field theories (cf. \cite{Stevenson:1983xb,Stevenson:1985zy,Soto:1986xy} for some early follow-up work on the subject).

Historically, concrete constructive evidence that classically unstable potentials can lead to positive definite Hamiltonian eigenspectra and perfectly well-defined unitary quantum mechanical time evolution was given by Bender and B\"ottcher \cite{Bender:1998ke}. They studied  non-Hermitian Hamiltonians of the form ${\cal H}=p^2-(i x)^\alpha$ with arbitrary $\alpha>2$ and showed how to use analytic continuation to calculate real and positive eigenenergies for the \textit{quantum} Hamiltonian ${\cal H}$ despite the fact that the \textit{classical problem} does not admit real and positive energies. It was later found by Jones and Mateo that the particular case $\alpha=4$ can be re-cast into an equivalent Hermitian eigenvalue problem \cite{Jones:2006qs}, with eigenvalue spectrum matching that from Bender and B\"ottcher \cite{Bender:1998ke}.

In the present work, I am going to formulate this finding as follows:

\textbf{In quantum mechanics, classically unstable potentials may be understood as analytic continuation of classically stable potentials into the unstable region, unless singular structures in the complex plane prohibit this analytic continuation.}

Based on this finding, I will push well-known results for field theories outside their classically trivial boundaries. As criterion to decide if a quantum field theory is well-defined I am proposing the following definition:

\textbf{A given quantum field theory is well-behaved in the continuum if all physical observables are well behaved. By contrast, classical intuition based on quantities that are not observables, such as in particular the value of the non-renormalization group invariant running coupling $\lambda_R(\bar\mu)$, should not be used.}

In a nutshell, if observables in a quantum field theory candidate come out well-behaved, this quantum field theory should be taken seriously even if it does not make sense classically. This alternative way of defining quantum field theory has surprising consequences.

\section{Explicit Calculations}

\subsection*{A toy model in 0d}

As a particular example, let me consider the simplest possible case of 0 dimensional field theory, with partition function possessing the integral representation
\be
\label{irep}
Z(\lambda)=\int_{-\infty}^\infty dx\, e^{-\lambda x^4}\,,\quad {\rm Re}(\lambda)>0\,.
\ee

Classically, the potential $V(x)=\lambda x^4$ is bounded only for ${\rm Re}(\lambda)>0$, which is what sets the limit on the above integral representation.

We can evaluate this integral for ${\rm Re}(\lambda)>0$, and find
\be
\label{two}
Z(\lambda)=2 \Gamma\left(\frac{5}{4}\right)\times \lambda^{-\frac{1}{4}}\,.
\ee
However, since the above result is valid in an open region for $\lambda$, we can analytically continue the result (\ref{two}) to values of $\lambda$ outside the domain of validity of the original integral representation (\ref{irep}). In our case, this is easy, because we can use the known analytic continuation of the root function. In particular, one finds for negative real $\lambda$
\be
Z(\lambda=-g)=(-1)^{-\frac{1}{4}} 2 \Gamma\left(\frac{5}{4}\right)\times g^{-\frac{1}{4}}\,,\quad g\in \mathbb{R}^+\,.
\ee
The analytically continued result for $Z(\lambda)$ to negative (real) $\lambda$ is not unique because of the four-sheeted nature of the quarter root. To obtain a unique result, additional information, such as a symmetry, is needed. For instance, if the additional information is that the partition function should be real and positive, the only possible result is
\be
\label{z4}
Z(\lambda=-g)=2 \Gamma\left(\frac{5}{4}\right)g^{-\frac{1}{4}}\frac{e^{\frac{i \pi}{4}}+e^{-\frac{i \pi}{4}}}{2}=\sqrt{2} \Gamma\left(\frac{5}{4}\right)g^{-\frac{1}{4}}\,,\quad g\in\mathbb{R}^+\,.
\ee

Far from being nonsensical, (\ref{z4}) is a perfectly well-behaved partition function for $\lambda<0$, despite the classically unbounded potential. 

The situation is completely analogous to well-studied functions in pure mathematics, such as the Riemann $\zeta$ function or the $\Gamma$ function, with integral representations defined by
\be
\zeta(s)=\frac{1}{\Gamma(s)}\int_0^\infty dx\, \frac{x^{s-1}}{e^{x}-1}\,, \quad \Gamma(s)=\int_0^\infty dx\, x^{s-1} e^{-s}\,,
\ee
for ${\rm Re}(s)>1$ and ${\rm Re}(s)>0$, respectively. The analytic continuation to negative real-valued arguments of these functions has been known for more than a century, in particular leading to $\zeta(-1)=-\frac{1}{12}$ and $\Gamma\left(-\frac{1}{2}\right)=-2 \sqrt{\pi}$. There is no controversy about evaluating $\zeta,\Gamma$ at negative real argument, so by analogy, neither should there be for analytic continuations such as (\ref{z4}).

\subsection*{The O(N) model in 4d}

The above toy model demonstrates the possibility of well-behaved results for the case of classically unbounded potentials. However, it is a toy model and not a bona-fide quantum field theory in four dimensions. Interacting four-dimensional field theories are in general extremely hard to solve, except if they possess a small parameter that allows non-perturbative expansions, such as many field components  \cite{Maldacena:1997re,Romatschke:2022jqg}. 

To be specific, let us consider the O(N) model defined by the Euclidean partition function
\be
Z=\int {\cal D}\vec{\phi}e^{-\int d^4x\left[\frac{1}{2}\vec{\phi}\left(-\partial_\mu\partial_\mu\right)\vec{\phi}+\frac{\lambda}{N}\left(\vec{\phi}^2\right)^2\right]}\,,
\ee
where $\vec{\phi}=\left(\phi_1,\phi_2,\ldots,\phi_N\right)$ is an N-component scalar field. Introducing an exact Hubbard-Stratonovic transformation with an auxiliary field $\zeta$, the large N limit of the O(N) model in any number of dimensions is given by (see Refs.~\cite{Romatschke:2019wxc,Romatschke:2019ybu,Romatschke:2021imm,Grable:2022swa,Romatschke:2022jqg,Romatschke:2023sce} for the detailed steps in between for various dimensions)
\be
\label{myz}
Z=\int_{-\infty}^\infty d\zeta_0 e^{-N\times {\rm vol}\times V_{\rm eff}(\sqrt{2 i \zeta_0})}\,,
\ee
with the effective potential in the large N limit given by
\be
V_{\rm eff}(m)=\frac{1}{2}\int \frac{d^4k}{(2\pi)^4}\ln\left(k^2+m^2\right)-\frac{m^4}{16 \lambda}\,.
\ee
In this form, the effective potential still suffers from UV-divergencies. The standard procedure to regulate divergences in high energy theory is dimensional regularization \cite{Workman:2022ynf}, though some researchers still prefer cut-off regularization despite it breaking Lorentz invariance of the theory. In either regularization scheme, the above integral is completely standard, and one finds in dimensional regularization (see Ref. \cite{Romatschke:2023sce} for cut-off regularization)
\be
V_{\rm eff}(m)=-\frac{m^4}{64\pi^2}\left(\frac{1}{\varepsilon}+\frac{4\pi^2}{\lambda}+\ln\frac{\bar \mu^2 e^{\frac{3}{2}}}{m^2}\right)\,,
\ee
where $\bar\mu$ is the $\overline{\rm MS}$ renormalization scale. The effective potential still needs to be renormalized, which in the present case is achieved by the non-perturbative renormalization condition
\be
\frac{1}{\varepsilon}+\frac{4\pi^2}{\lambda}=\frac{4\pi^2}{\lambda_R(\bar\mu)}\,,
\ee
with the exact large N running coupling $\lambda_R(\bar\mu)$ having $\beta$ function
\be
\beta\equiv \frac{\partial \lambda_R(\bar\mu)}{\partial \ln \bar\mu^2}=\frac{\lambda_R^2(\bar\mu)}{4\pi^2}\,.
\ee
The large N exact $\beta$-function is uniformly positive for all real $\lambda_R$. Integrating the $\beta$ function, one obtains for the explicit large N exact running coupling
\be
\label{exact}
\lambda_R(\bar\mu)=\frac{4\pi^2}{\ln \frac{\Lambda_{\overline{\rm MS}}^2}{\bar\mu^2}}\,,
\ee
where $\Lambda_{\overline{\rm MS}}$ is the $\Lambda$ parameter of the O(N) model, in complete analogy to what is done in QCD \cite{Workman:2022ynf}.

For small values of $\bar\mu\ll \Lambda_{\overline{\rm MS}}$, the running coupling is positive, allowing a simple and intuitive classical interpretation of the theory. This is the regime in which scalar field theory is usually employed, as a cut-off (effective) theory for scales $\bar\mu\ll \Lambda_{\overline{\rm MS}}$.

Increasing $\bar\mu$, one finds that the running coupling increases and finally diverges at $\bar\mu=\Lambda_{\overline{\rm MS}}$, which is often referred to as the Landau pole of the theory. Again, classical intuition fails near the Landau pole, even though several example in the literature exist where observables remain well-defined and finite despite the divergent coupling, e.g. \cite{Aharony:1999ti,Policastro:2001yc,gurarie2007resonantly,Romatschke:2021imm,Grable:2022swa,Lawrence:2022vwa}. Common lore also has it that near the Landau pole, all higher dimension operators turn on, rendering the theory incalculable. This is a myth, as shown in Ref.~\cite{Romatschke:2022llf}.

Beyond the Landau pole, $\lambda_{R}(\bar\mu)$ remains well-defined, increasing, but negative for $\bar\mu>\Lambda_{\overline{\rm MS}}$, straining classical interpretation. For asymptotically high energies $\bar\mu\rightarrow \infty$, $\lambda_R(\bar\mu)$ approaches zero (albeit from below), which demonstrates that the O(N) model is an example of an asymptotically free theory. However, from the point of view of analytic continuation, nothing particularly remarkable is happening. It is key, however, to note that the non-positivity of $\lambda_R(\bar\mu)$ in the UV is exploiting precisely the loophole in the proofs of asymptotic freedom and quantum triviality \cite{Coleman:1973sx,Aizenman:2019yuo}. This is not an engineered setup, it follows naturally from solving the O(N) model non-perturbatively in the large N limit, and has been known for decades \cite{Abbott:1975bn,Linde:1976qh}. 

According to the criterion outlined in the introduction, it is necessary to calculate observables in order to decide if the theory is well-behaved. Fortunately, one can easily calculate observables in the large N limit of the O(N) model. The observable that is most easily accessible is the value of the partition function itself, which in the large N limit is given exactly from the saddle point of the integral (\ref{myz}). One finds for the free energy density \cite{Romatschke:2022jqg}
\be
   {\cal F}=-\frac{\ln Z}{\rm vol}=N V_{\rm eff}(m)\,,
   \ee
   with $m$ given by $\frac{dV_{\rm eff}(m)}{dm}=0$\,.

   Inserting the explicit form of the large N exact running coupling (\ref{exact}) into the renormalized expression for $V_{\rm eff}$ one finds
   \be
   V_{\rm eff}(m)=-\frac{m^4}{64\pi^2}\ln \frac{\Lambda_{\overline{\rm MS}}^2 e^{\frac{3}{2}}}{m^2}\,.
   \ee

   One finds two saddles for the partition function: $m=0$, and $m=\sqrt{e}\Lambda_{\overline{\rm MS}}$. The free energy density for these saddles is
   \be
   \label{fres}
   {\cal F}_{m=0}=0\,,\quad {\cal F}_{m=\sqrt{e}\Lambda_{\overline{\rm MS}}}=-\frac{N e^2 \Lambda_{\overline{\rm MS}}^4}{128\pi^2}\,.
   \ee
   Since the free energy is an observable, it cannot depend on the fictitious renormalization scale $\bar\mu$, and it is gratifying to see that this is indeed the case for (\ref{fres}). Both results for the free energy are well-behaved, showing no sign of any pathologies that a simplistic classical interpretation of the potential would have perhaps suggested. This is no accident: the value of the running coupling, with its fictitious renormalization scale dependence, cannot appear on its own in any observable, and indeed it does not for the free energy. Put differently: it is irrelevant that the running coupling diverges at the Landau pole or that it becomes negative in the UV, because the free energy is not directly sensitive to this fictitious renormalization-scale dependent quantity.

   The value of the free energy is important, however. Basic thermodynamics tells us that in the presence of two phases, the phase with the lower free energy is thermodynamically preferred. This means that the saddle point solution $m=0$ is thermodynamically unstable with respect to decay to the thermodynamically preferred saddle $m=\sqrt{e}\Lambda_{\overline{\rm MS}}$, something which confused early researchers \cite{Coleman:1974jh} but was clarified soon afterwards \cite{Abbott:1975bn,Linde:1976qh}.

   Besides the free energy, another observable is the pole mass of the vector $\vec{\phi}$, which is given by the value of the saddle (see e.g. Ref.~\cite{Romatschke:2022jqg} for details on the calculation). For the thermodynamically preferred phase, the large N exact result for the pole mass
   \be
   m=\sqrt{e}\Lambda_{\overline{\rm MS}}\,,
   \ee
   is again renormalization-scale independent, well-behaved and free from any pathologies.

   One might worry that pathologies only show up when considering scattering, which requires consistently including $\frac{1}{N}$ corrections into the calculation. Fortunately, this is not hard to do, and one finds for the cross section for example in the s-channel
   \be
   \label{complete}
   \sigma(E)=\frac{(4 \pi)^3}{N^2 E^2 \left|1-2 \sqrt{1-\frac{4 m^2}{E^2+i 0^+}}{\rm atanh}\frac{1}{\sqrt{1-\frac{4m^2}{E^2+i 0^+}}}\right|^2}
   \ee
   to NLO in large N. Again, the cross section is renormalization-scale independent, well-behaved and free from any pathologies. The only curious feature of the cross section is the presence of a stable scalar bound state with a mass of
   \be
   m_2\simeq 1.84 m\,,
   \ee
   that was again already found a long time ago \cite{Abbott:1975bn}. Consistent incorporation of NNLO corrections in the large N expansion are expected to imbue this scalar bound state with a finite width, in complete analogy to how muonium obtains a finite width in perturbative QED calculations \cite{Peskin:1995ev}. The result for the scalar mass is renormalization-scale independent, well-behaved and free of pathologies.

   It should be stressed that calculating the cross-section in perturbation theory one encounters divergencies from the Landau pole at every single order in perturbation theory, cf. the discussion in lecture 3 in \cite{Romatschke:2023ztk}. However, expanding out (\ref{complete}) in perturbation theory one finds that it corresponds to the sum over an infinite number of perturbative ``bubble diagrams'', rendering the end-result finite and insensitive to the Landau pole. 

Non-perturbative evaluation for the O(N) model at finite N can be done by discretizing negative coupling field theory on a space-time lattice. The corresponding lattice action appears to be non-polynomial \cite{Romatschke:2023sce,Weller:2023jhc}, yet amenable to numerical integration e.g. for N=1\cite{Romatschke:2023sce,Romatschke:2023fax}. Further numerical work for negative coupling scalar field theory on the lattice is required to study if the large N results carry over to N=1,2.
   
   \section{Summary and Conclusions}

   In this work, I propose to use observables instead of classical intuition in order to decide whether or not a quantum field theory is sensible in the continuum. I show that --- based on this criterion --- the O(N) model in four dimensions in the large N limit is a sensible interacting quantum field theory exhibiting asymptotic freedom, despite (or perhaps because) it possesses a Landau pole and non-positive running coupling in the UV. This property (non-positive coupling) is precisely the loophole in no-go proofs for four-dimensional scalars  in Refs.~\cite{Coleman:1973sx,Aizenman:2019yuo}, rendering both proofs ineffective for the four-dimensional O(N) model in the large N limit. Further work needs to be done in order to decide if theories with a finite (small) number of scalars, such as the N=4 Higgs field in the Standard Model, are also non-trivial and asymptotically free quantum field theories.

   \section{Acknowledgements}

   I thank the particle physics groups at the University of Ljubljana, the Technical University of Vienna and the University of Washington for their hospitality and helpful discussions on this topic. Also, I thank Michael Aizenman for encouraging remarks concerning the existence of negative coupling field theory, and Seth Koren for helpful discussions on fleshing out the phenomenological applications to the Standard Model Higgs sector. This work was supported by the Department of Energy, DOE award No DE-SC0017905.

\bibliography{PT}
\end{document}